\documentclass[letterpaper]{IEEEtran}

\usepackage{cite}                        
\usepackage[pdftex]{graphicx}                      
\usepackage[cmex10]{amsmath}      		  
\usepackage{array}                  
\usepackage{mdwmath}    
\usepackage{mdwtab}                         	
\usepackage{amssymb}     	

\usepackage{url}

\begin{document}

\markboth{This article has been accepted for publication. DOI: 10.1109/LAWP.2015.2458352, IEEE Antennas and Wireless Propagation Letters}{}

\title{
	\begin{center}
	A New Approach to the Link Budget Concept \\
	for an OAM Communication Link
	\end{center}
	}
	
\author{Andrea Cagliero, Assunta De Vita, Rossella Gaffoglio and Bruno Sacco%
\thanks{A. Cagliero and R. Gaffoglio are with the Department of Physics, University of Torino, I-10125 Torino, Italy (e-mail: andrea.cagliero@unito.it; rossella.gaffoglio@unito.it).}
\thanks{A. De Vita and B. Sacco are with the Centre for Research and Technological Innovation, RAI Radiotelevisione Italiana, I-10128 Torino, Italy (e-mail: assunta.devita@rai.it; bruno.sacco@rai.it).}
\thanks{{\textbf{1536-1225 (c) 2015 IEEE. Personal use is permitted, but republication$/$redistribution requires IEEE permission. See}}
http:$//$www.ieee.org$/$publications$\_$standards$/$publications$/$rights$/$index.html {\textbf{for more information.}}}}


\maketitle

\begin{abstract}

Following on from the increasing interest for electromagnetic waves carrying Orbital Angular Momentum (OAM), different configurations of antenna systems able to generate such beams have been proposed. However, in our opinion, a traditional radiation pattern approach does not provide the right picture of an OAM-based communication link. For this reason we propose a new general concept, the ``OAM-link pattern'', which takes into account the peculiar phase structure characterizing these waves. Focusing on OAM transmissions between antenna arrays, we introduce a formula for the link budget evaluation which describes the whole communication link and directly leads to a ``classically shaped'' main lobe pattern for a proper rephased reception in the case of uniform circular arrays.

\end{abstract}


\IEEEpeerreviewmaketitle

\section{Introduction}

\IEEEPARstart{I}{t} has been known for a long time that electromagnetic waves can carry angular momentum, but only in 1992 it was found theoretically by L. Allen \emph{et al.} \cite{AllenEtAl1992} that some types of beams possess an Orbital Angular Momentum (OAM) of $\ell\hbar$ per photon, where $\ell\in\mathbb{Z}$ is the so-called azimuthal index. These beams are vortex waves characterized by a phase dislocation of the form $e^{i\ell\phi}$, where $\phi$ is the azimuthal angle, which determines helical wavefronts and a doughnut-shaped beam profile with a zero on-axis intensity for $\ell\neq 0$. Another feature of OAM beams is the orthogonality among modes with different $\ell$ index, which, in principle, allows to convey independent information channels on a single frequency. 

Although these twisted waves are extensively used in modern optical experiments \cite{GibsonEtAl2004,WangEtAl2012}, only recently F. Tamburini \emph{et al.} \cite{TamburiniEtAl2012a} proposed the idea of employing them in free-space radio communications. Such a possibility is claimed to offer a solution to the current problem of radio band congestion and paves the way to a huge literature on the subject, not exempt from harsh criticism \cite{TamagnoneEtAl2012, TamburiniEtAl2012b, TamagnoneEtAl2013, EdfordsJohansson2012}. In fact, the potential use of such degrees of freedom in the radio domain, by means of present technologies, seems to reveal intrinsic limitations due primarily to the peculiar topology of the OAM profiles, characterized by a central dark region that increases with distance \cite{TamagnoneEtAl2013}. This problem becomes significant when considering far-field radio communications, where the divergence of the beams is not negligible and only a small fraction of the intensity profile can actually be intercepted by the receiving antenna. 

In order to get insights on the effective relevance and usability of these vortex waves, it is fundamental to take into consideration the whole OAM-based communication link, rather than focusing on the radiation pattern associated with the transmitted field. This assumption, which has general validity, becomes especially relevant in a radio communication scenario, where it naturally leads to reconsider the concept of ``link budget''. A pioneering work in this sense was carried out by D. K. Nguyen \emph{et al.} \cite{Nguyen_Eucap2014} for the case of vortex waves generated by arrays of point-like isotropic radiators. 

In this letter we introduce the general concept of ``OAM-link pattern'', which takes into account not only the physical features of the transmitted beams, as a radiation pattern merely does, but also the proper way to receive them according to the helical structure characterizing their wavefronts. We show that, when applied to OAM transmissions between antenna arrays, the OAM-link pattern proves helpful for a correct link budget estimate. Our general considerations are finally implemented in the case of circular arrays composed by half-wave dipoles.

\section{OAM beams generation}

Vortex waves are well-known in paraxial optics, where they are generally associated to Laguerre-Gaussian beams, Bessel-Gaussian beams and other solutions to the scalar paraxial wave equation \cite{GoriEtAl1987, AllenEtAl1992, Gutierrez-Vega2005}. OAM solutions to the exact scalar and vector Helmholtz equations are also known \cite{Durnin1987, VectorBessel}, but they usually play a minor role due to the fact that they carry infinite energy and therefore are not physically realizable.

A great number of different methods have been implemented to successfully transform an untwisted beam into a wave carrying OAM, leading to good approximations of the ideal electromagnetic modes. Among the several techniques able to produce vortex waves, the simplest methods, more commonly employed in Optics, require the use of computer-generated holograms, cylindrical lenses \cite{PadgettAllen2000} or special devices called Spiral Phase Plates (SPPs), which are constructed by shaping a piece of transparent material in such a way as to have a gradually increasing, spiraling thickness with a refractive index different from that of air \cite{ Berry2004}. 
However, in the domain of the radio frequencies, two other methods are mainly used to approach the production of OAM radiation. The first one requires the reflection of an untwisted radio wavefront on a properly shaped surface. This is the case of the twisted parabolic antenna, whose dish is transformed into a vortex reflector by progressively elevating, from the original shape, its surface, according to the different values of the azimuthal angle $\phi$ \cite{Trinder}. The other procedure consists in a direct action on the current supplies relative to the elements of an antenna array, so that the beam produced presents a twisted structure. According to the so-called array-synthesis method, it is possible to obtain, by means of a linear inverse problem, the input currents with known location that best approximate a given ideal electromagnetic mode. Hence, starting from an array with identical radiating elements, arranged according to a fixed layout, this method can be applied to achieve a good reproduction of an ideal OAM mode (e.g. a Laguerre-Gaussian beam).

\section{OAM-link pattern}

When considering a radio-wave transmission, one of the most relevant aspects to take into account is the so-called antenna pattern or antenna gain $G(\vartheta,\varphi)$, which represents the modulus of the radial component of the normalized Poynting vector associated with the radiated far field as a function of the polar angles \cite{Orfanidis}. This pattern plays a fundamental role for the link budget estimate and can be used to compute the transmission equation by means of a straightforward rule. As it is widely known, given a transmitting system $T$, a receiving system $R$ and the corresponding antenna patterns $G_T(\vartheta_t,\varphi_t)$ and $G_R(\vartheta_r,\varphi_r)$, the ratio between the received and the transmitted powers in free-space can be simply estimated via the Friis formula:
\begin{equation}
\frac{P_R}{P_T}= \left( \frac{\lambda}{4\pi d} \right)^2 G_T G_R,
\label{eq_Friis}
\end{equation}
where $\lambda$ is the operating wavelength, $d$ represents the link distance and $G_T$, $G_R$ correspond to the gains along the $T$-$R$ direction; antennas polarization is neglected for simplicity. 

Usually, in standard radio communications, the antenna pointing direction is that of maximum gain (\emph{main lobe}) and the ratio $P_R/P_T$ can be easily derived from (\ref{eq_Friis}) once the main lobe gains of the transmitting and receiving systems are provided. In the OAM case, for $\ell\ne0$, the antenna patterns are known to exhibit the characteristic on-axis minimum \cite{Mohammadi2010}; since the main lobe is located at a certain $\vartheta\ne0$, it would seem that the best antenna pointing should be shifted accordingly. However, as remarked by D. K. Nguyen \emph{et al.}, the attempts to apply the above described rule for the link budget estimate to the OAM case inevitably lead to inconsistent results \cite{Nguyen_Eucap2014}. Indeed, the relevant information of an OAM beam is represented by the peculiar structure of its wavefront, which degrades far away from the on-axis phase singularity and it is not accessible in its close vicinity because of the central intensity drop. This information is lost in the implicit assumptions of (\ref{eq_Friis}), which relies on the concepts of point-like antennas and far-field limit. From this point of view, it is misleading to speak of antenna pattern within the context of an OAM-based transmission. What we need is a new formulation for the link budget, which takes into account the mode-matching between wavefront and antenna, exploiting the information that cannot be inferred from the usual antenna pattern. The proposed link budget defines, as a function of the azimuth and elevation angles of one of the two systems, what we might call an ``OAM-link pattern''. Aside from its general validity, in this letter we show how this concept finds a straightforward implementation in the case of OAM transmissions between antenna arrays.

\section{Analytical formulation}

As already mentioned, a general way to produce an OAM beam in the radio domain is to use an array of $N_{T}$ identical radiators, fed with a set of currents provided by the synthesis of an ideal vortex mode. The reception of such a beam can be performed using another array placed in front of the transmitting one, at a fixed distance, perpendicularly to the line connecting their centers. Considering the generation of an OAM beam with an azimuthal index $\ell_T$, the circuit voltage induced on a $p$th receiving element can be written as the dot product between its effective height and the incoming electric field \cite{Orfanidis}:
\begin{equation} \label{eq:res3}
V^{\ell_T}_p=ik\eta\: \sum_{n=1}^{N_{T}}\:\frac{e^{-ik r_{np}}}{4\pi r_{np}}\: I^{\ell_T}_n \:\mathbf{h}^{\substack{\text{{\tiny{T}}}}}_{np}\cdot\mathbf{h}^{\substack{\text{\tiny{R}}}}_{pn},
\end{equation}
being $k=2\pi/\lambda$ the modulus of the wave vector, $\eta$ the vacuum impedance and $I^{\ell_T}_n=(I_0/\sqrt{N_T})\:\xi^{\ell_T}_n$ the input current to the $n$th antenna terminals, where $\xi^{\ell_T}_n$ is the coefficient arising from the synthesis of the $\ell_T$ mode considered, while $I_0$ is a constant current term associated to the total input power through the following relation:
\begin{equation} \label{eq:res3bis}
P_{in}=\frac{1}{2}\:I^2_0 R\:\frac{1}{N_T}\:\sum_{n=1}^{N_T}\left|\xi^{\ell_T}_n\right|^2=\frac{1}{2}\:I^2_0 R,
\end{equation}
where $R$ is the resistance of the radiators and the orthonormality condition is assumed for the synthesis coefficients. Moreover, $\mathbf{r}_{np}$ is a vector that points from the $n$th transmitting antenna to the $p$th receiving one, while:
\begin{equation} \label{eq:res4}
\mathbf{h}^{\substack{\text{{\tiny{T}}}}}_{np}=\mathbf{h}^{\substack{\text{{\tiny{T}}}}}(\vartheta_{np},\varphi_{np}) \ \ \mbox{and} \ \ \mathbf{h}^{\substack{\text{{\tiny{R}}}}}_{pn}=\mathbf{h}^{\substack{\text{{\tiny{R}}}}}(\vartheta_{pn},\varphi_{pn})
\end{equation}
are the effective heights relative to each reciprocal link, evaluated at the angular coordinates $(\vartheta,\varphi)$ defining the direction from the $n$th transmitting element to the $p$th receiving one, and vice versa, according to the corresponding reference frame. Mutual couplings between antennas are not taken into account.

The total voltage induced on the whole receiving array can be easily obtained summing (\ref{eq:res3}) over all its elements. However, due to the peculiar phase pattern associated with an OAM beam with $\ell\neq 0$, this standard reading of the received signals inevitably leads to a reciprocal cancellation, giving rise to a null on-axis power transfer response. The above considerations suggest that the phase structure of an OAM wave must be taken into account by introducing a phase-weighting at the receiver; this naturally leads to the following definition for the total received power \cite{Orfanidis}:
\begin{equation} \label{eq:res5}
P^{\ell_{T,R}}_{out}=\frac{1}{8R}\left|V^{\ell_{T,R}}_{out}\right|^2=\frac{1}{8R}\left|\frac{1}{\sqrt{N_R}}\:\sum^{N_{R}}_{p=1}\xi^{-\ell_R}_p\:V^{\ell_T}_p\right|^2,
\end{equation}
where $V^{\ell_{T,R}}_{out}$ is the total rephased output voltage and $\xi^{-\ell_R}_p$ is the coefficient assigned to the $p$th receiving element and corresponding to the synthesis of an OAM mode with index $-\ell_R$ (a minus sign is needed, being the two identical arrays placed one in front of the other). In (\ref{eq:res5}) only a receiving configuration with $\ell_R=\ell_T$ allows the proper rephasing, and thus the correct reading, of the incident vortex wave with the azimuthal index $\ell_T$. Hereinafter we will show that this behaviour directly follows from the orthogonality property of the OAM modes.    

Substituting (\ref{eq:res3}) in (\ref{eq:res5}) and taking into account (\ref{eq:res3bis}), the following relation is obtained:
\begin{equation} \label{eq:res5bis}
\frac{P^{\ell_{T,R}}_{out}}{P_{in}}=\left|\frac{ik\eta}{2R}\frac{1}{\sqrt{N_TN_R}}\sum_{p=1}^{N_R}\sum_{n=1}^{N_T}\frac{e^{-ik r_{np}}}{4\pi r_{np}} \xi^{\ell_T}_n\xi^{-\ell_R}_p\mathbf{h}^{\substack{\text{{\tiny{T}}}}}_{np}\cdot\mathbf{h}^{\substack{\text{\tiny{R}}}}_{pn}\right|^2
\end{equation}
which provides a link budget evaluation for the OAM channel considered. 

In order to probe this ratio as the orientation of the receiving array varies, we can rigidly rotate its entire structure around its center by an elevation angle $\alpha$ and an azimuth angle $\beta$. The result is the sought-for OAM-link pattern, parameterized with the angles $\alpha$ and $\beta$ and depending on the type of antenna used, the current supplies, the geometry of the arrays and the receiving configuration. 

\begin{figure}[!t]
\centering
\includegraphics[scale=0.3]{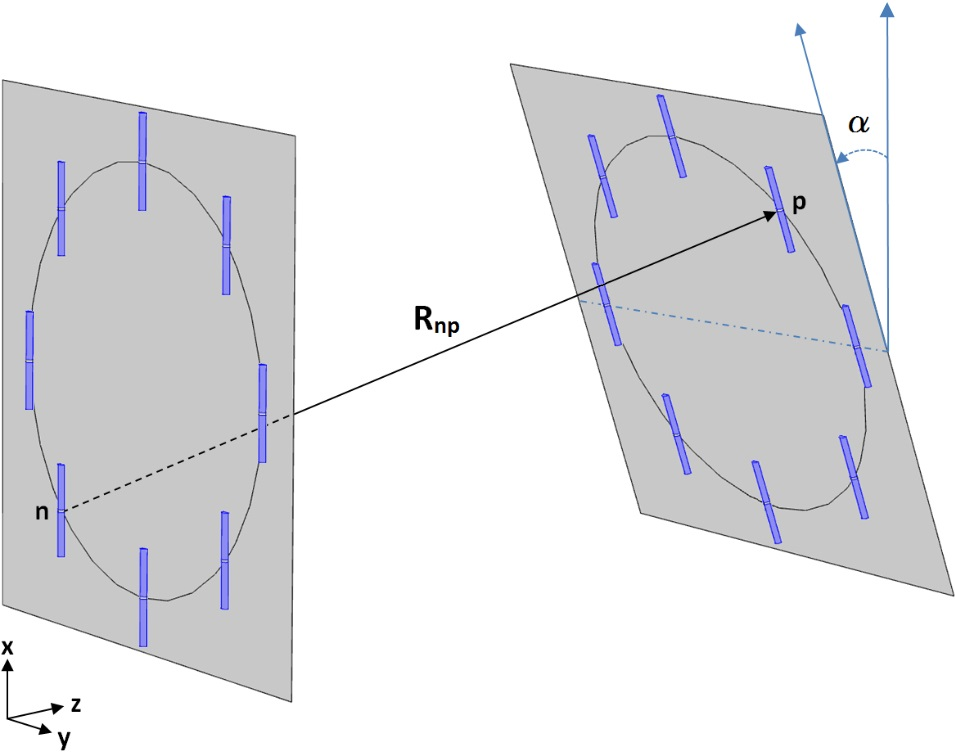}
\caption{Schematic representation of two facing uniform circular arrays composed by eight elements.}
\label{fig_arrayUCAdefinitivo}
\end{figure}

\section{OAM-link pattern for uniform circular arrays}
The discussion about the link budget previously developed for two facing arrays takes into account each individual link between a transmitting antenna and a receiving one, considering their respective gains and geometrical locations. Thanks to its general validity, we can apply our formulation to an OAM transmission between two circular arrays with the same number of elements (as an example, see Fig. \ref{fig_arrayUCAdefinitivo}). In this case $N$ transmitting antennas are fed with the same current amplitude $I_0/\sqrt{N}$, but successively delayed such that after a full turn around the antenna array axis the phase experiences a variation of $2\pi\ell_T$, being $\ell_T$ the desired OAM index \cite{Mohammadi2010}. The input currents can thus be written as:
\begin{equation} \label{eq:th1}
I^{\ell_T}_n=\frac{I_0}{\sqrt{N}}\:\Phi_{\ell_T}^n\equiv \frac{I_0}{\sqrt{N}}\:\exp\left[i\ell_T\:2\pi\left(\frac{n-1}{N}\right)\right]
\end{equation}
for $n=1,...,N$; whereas the relation that determines how many $\ell$ modes the array can generate is: $-N/2<\ell<N/2$ and for the discrete set $\left\{\Phi^n_{\ell}\right\}$ the following orthogonality condition holds:
\begin{equation} \label{eq_orthogonality2}
\frac{1}{N}\sum_{n=1}^{N}\left(\Phi_{\ell_{R}}^{n}\right)^{*}\Phi_{\ell_{T}}^{n}=\delta_{\ell_{T},\ell_{R}}.
\end{equation}
In this context, according to (\ref{eq:th1}), the total received power (\ref{eq:res5}) can be rewritten as:
\begin{equation} \label{eq:res6bis}
P^{\ell_{T,R}}_{out}=\frac{1}{8R}\left|\frac{1}{\sqrt{N}}\sum^{N}_{p=1}\exp\left[-i\ell_R\:2\pi\left(\frac{p-1}{N}\right)\right]V^{\ell_T}_p\right|^2,
\end{equation}
where the OAM orthogonality is maximized when both the transmitting and the receiving arrays are perpendicular to the line connecting their centers.

Focusing on a realistic example, expression (\ref{eq:res6bis}) can be evaluated for two uniform circular arrays of identical half-wave dipoles by simply inserting in (\ref{eq:res3}) the proper effective height reported below for a couple of facing dipoles $(n,p)$:
\begin{equation} \label{eq:res7}
\mathbf{h}(\vartheta_{np},\varphi_{np})=\frac{1}{k}\:\frac{2}{\sin\vartheta_{np}}\:\cos\left(\frac{\pi}{2}\:\cos\vartheta_{np}\right)\:\mathbf{\hat{\vartheta}}_{np}.
\end{equation}
Expression (\ref{eq:res7}) can be easily derived following the approach presented in \cite{Orfanidis}. Choosing the $x$-axis to be along the direction of the antennas and assuming $\hat{z}$ as the axis of propagation perpendicular to the planes where the circular arrays lie, the polarization versor acquires the form: $\mathbf{\hat{\vartheta}}_{np}=-\mathbf{\hat{x}}\:\sin\vartheta_{np}+\mathbf{\hat{y}}\:\cos\vartheta_{np}\:\cos\varphi_{np}+\mathbf{\hat{z}}\:\cos\vartheta_{np}\:\sin\varphi_{np}$. Being $\mathbf{r}_{np}$ the vector connecting the dipole $n$ to $p$, $\vartheta_{np}$ is the angle that $\mathbf{r}_{np}$ forms with the $x$-axis, while $\varphi_{np}$ is the angle between the $y$-axis and the projection of $\mathbf{r}_{np}$ onto the {\itshape y-z} plane. 

\begin{figure}[!b]
\centering
\includegraphics[scale=0.6]{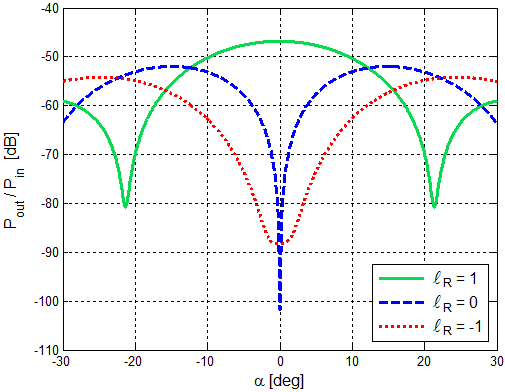}
\caption{Total received power normalized to the total radiated power as a function of the rotation angle $\alpha$ relative to the receiving array. The OAM beam transmitted has an azimuthal index $\ell_T=1$, while $\ell_R$ refers to three different rephasing configurations.}
\label{fig_linearplot}
\end{figure}

\begin{figure}[!t]
\centering
\includegraphics[scale=0.75]{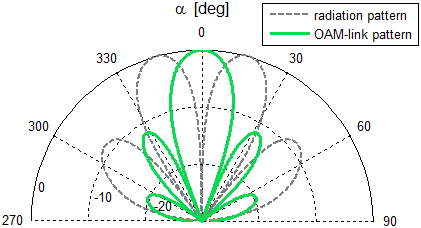}
\caption{{\itshape{Solid line}}: OAM-link pattern (total received power in {\itshape dB} as a function of the rotation angle $\alpha$ for the receiving configuration $\ell_R=\ell_T=1$); {\itshape{dashed line}}: conventional radiation pattern for $\ell_{T}=1$. Both the curves are normalized to their respective maxima. It should be noted that the plotted quantities represent two distinct concepts from a physical point of view.}
\label{fig_polarplot}
\end{figure}

In the plots here presented, obtained with the {\scshape Matlab}\textsuperscript\textregistered $\!$ software \cite{MATLAB2014}, we considered a couple of identical circular arrays, as those shown in Fig. \ref{fig_arrayUCAdefinitivo}, with radius $R=1.5\:m$, made of eight equally spaced half-wave dipoles and separated by a link distance of 40 meters. The transmitting radiators are phased according to (\ref{eq:th1}) in order to produce an OAM mode with $\ell_T=1$ at a frequency of about $205\:MHz$ (being $\lambda=1.46\:m$). Fig. \ref{fig_linearplot} shows the total received power {\itshape vs} the array rotation angle $\alpha$ for three different receiving phase configurations: $\ell_R=1,0,-1$. As expected, at $\alpha=0$ the orthogonality of the OAM modes at its maximum together with the polarization matching of the dipoles allow for a significant reception of the signal only when the rephasing configuration is $\ell_R=\ell_T=1$ (i.e. the solid green curve) presenting an on-axis power maximum in correspondence of the conventional radiation pattern null. This remarkable behaviour is displayed more clearly in the polar plot shown in Fig. \ref{fig_polarplot}.

The absence of an on-axis zero in the received power relative to the destructive configuration (i.e. the dotted red curve in Fig. \ref{fig_linearplot}) is due to the directivity of the considered array elements.

It should be noted that the far-field evolution of this pattern with respect to the link distance $d$ is found to be in accordance with the well known decay $d^{-2\ell-2}$, which directly follows from the topology and divergence of our synthesized OAM profiles. 

Finally, we stress that our formulation can be extended, in principle, to the case of aperture or reflector systems by means of the reciprocity theorem; this acquires particular relevance in the optical regime, where both the production and the detection of OAM beams are often carried out by means of continuous spiral phase plates.

\section{Conclusions}

In this letter we have introduced the concept of ``OAM-link pattern'', which leads to a formal description of an OAM-based transmission link. We emphasized its particular relevance in a radio communication context, where it provides the correct way to implement an OAM link budget between antenna arrays, and we supported our suggestion by means of a rigorous calculation in the case of uniform circular arrays. As a consequence of the orthogonality of the OAM modes, which is highly sensitive to misalignment, a classic on-axis main lobe actually arises in this new angular pattern, once considered the proper rephased reception.

\section{Acknowledgments}
We wish to thank Prof. Paolo Gambino and Prof. Roberto Tateo, from the University of Torino, Department of Physics, and Prof. Giuseppe Vecchi, from the Polytechnic University of Torino, for sharing their knowledge with us by means of helpful comments and fruitful discussions.

\end{document}